\documentclass{article}
\usepackage{axodraw}
\begin{document}
\begin{center}
{{\bf \Large Research News---\\  Recent Advances in
the Study of B-meson systems
}}

\bigskip

Gauhar Abbas$^a$,\\
B. Ananthanarayan$^a$,\\ 
Kriti Ashok$^b$,\\
S. Uma Sankar$^c$

\bigskip
{\small
$^a$ Centre for High Energy Physics, Indian Institute of Science,
Bangalore 560 012\\
$^b$ St. Stephen's College, Delhi 110 007\\
$^c$ Department of Physics, Indian Institute of Technology, Bombay 400 076
\\
}
\medskip
\end{center}


\medskip
\begin{abstract}
An accessible review of recent discoveries concerning $B$-meson systems 
including those from Belle and BaBar on direct CP asymmetries, and the 
measurement of the width difference of $B_s$-meson system at Fermilab 
is presented.
\end{abstract}

\bigskip

A recent publication~\cite{BELLE1}
from the Belle experiment at the KEK B-factory
in Japan, in the journal {\it Nature} reports the
observation of a significant difference in the measurements of
the `direct charge-parity (CP) violation between charged and
neutral B meson decays'.  The experiment looks at the relative difference
in the rates from the decays $B^\pm\to K^\pm \pi^0$ and the
corresponding rates from their neutral counterparts.  Although
these differences are expected to be similar, the experiment finds
them to be significantly different.  
Somewhat earlier, the BaBar experiment at the Stanford Linear
Accelerator Center (SLAC) B-factory, USA also carried out significant 
related studies~\cite{BABAR1,BABAR2}.
The same issue of the journal {\it Nature} carries
an article by the well-known theorist Michael E. Peskin~\cite{Peskin}
which explains the significance of this measurement and 
presents a detailed discussion on its implications to theory
as well as experiment.

The measurements here pertain to a phenomenon known as `direct' CP
violation in the B-meson system, where C stands for charge conjugation,
a symmetry that exchanges particles for their anti-particles and
vice versa, and P for parity, or mirror symmetry.  
CP violation is of central importance in the field of elementary particle
physics and also in cosmology.  For instance, the famous Russian
physicist, Andrei Sakharov proposed that CP violation is one of the
three required phenomena which underpin our
understanding as to why there is an excess of matter over antimatter in
the Universe (with a net baryon number), 
if the Universe began from an initial state where
it was matter-antimatter symmetric (with zero baryon number). 
The other two phenomena are baryon number violating interactions
(not yet observed) and departure from thermal equilibrium.  
Baryon number violating interactions occur naturally
in grand unified models of elementary particle interactions, into
which the `standard model', viz., the theory of electroweak and strong
interactions, can be embedded.  The standard model as we understand
it today, itself has the seeds of CP violation, although these may be 
insufficient to generate the observed baryon number asymmetry.
A prior publication in this journal~\cite{Panchapakesan} gives
a lucid introduction to these ideas.
In the following, we explain why the newly observed differences in 
the direct CP violation of charged and neutral B mesons, may be of
great importance.

We recall here that mesons are composite particles made up of 
a quark and an anti-quark
which could be of different `flavours'.  Quarks, which are believed
to be elementary particles with strong interactions, come
in six different flavours, namely, $u-,\, d-,\, s-,\, c-,\, b-$ and $t$.   
$u-,\, c-$ and $t-$ quarks have electric charge $2/3$ and $d-,\, s-$ and $b-$
quarks have electric charge $-1/3$, in units of proton charge.
Of the B-mesons observed by BELLE, charged ones are made up of the 
flavours $B^+(=\overline{b} u)$ and $B^-(=\overline{u} b)$ and
the neutral ones are of flavours $B^0 (\overline{b} d)$, 
$\overline{B}^0 (\overline{d} b)$. If in these, the b- quark were 
to be replaced by the s- quark, it would represent the K-meson system.
In the six quark model, there is a subtle phenomenon that occurs:
$d-,\, s-$ and $b-$quarks (all of which have the same electric charge)
that would be stable in the absence of the weak interactions, 
get `mixed' amongst one another in accordance with the laws of 
quantum mechanics, when the weak interactions are present. Such mixing
would be governed by the action of a most general $3\times 3$ unitary 
matrix with complex entries. However, simple redefinitions of phases 
of the fields that describe each of the quark flavours can make most 
of the entries in this matrix real. It was shown by Kobayashi and 
Maskawa~\cite{Kobayashi:1973fv}
that the resulting matrix, after the phase redefinitions, 
would be parametrized by three angles that would enter the matrix 
through sines and cosines,
and one parameter that would enter the matrix as a `phase' $e^{i\delta}$.
This latter, which represents the unalterable complex nature of the
quark mixing, would lead to CP violation.  
The bold prediction of Kobayashi and Maskawa is that every instance of 
CP violation 
can be consistently fitted to the value of this unique parameter $\delta$.
The ideas of Kobayashi and Maskawa were remarkably prescient as they
were presented when the three heaviest quark flavours were actually
undiscovered.  However, it showed that the presence of heavy quarks
was somehow needed to account for CP violation.  
They also predicted that CP violation is expected
to be more dominant in the B-meson system than in the K-meson system.

Indeed, the Nobel prize winning discovery of CP violation
in 1964 was in the neutral K-meson system through
so-called `indirect' CP violation which proceeds through `mixing' of
a neutral K-meson $(\bar{s} d)$ with its anti-particle $s \bar{d}$  
(A neutral particle {\bf is not} its own anti-particle, if it has
`charges' other than electric charge). This mixing, whose origin can be
traced to the `mixing' of the quarks of the same charge mentioned above,
leads to the production of one long lived ($K_L$) and
one short lived ($K_S$) state.  CP conservation would have implied that
$K_L$ would never decay into two-pions.  The fact that it does
is the hallmark of CP violation in this indirect manifestation of
CP violation.
The discovery of `direct' CP violation even in the K-meson system
had to wait until the year 1999, as it is a very fine effect in this
system.  This comes from the observation of a difference in the
rates for the decay of $K_L$ into $\pi^+ \pi^-$ and $\pi^0\pi^0$
pairs by the KTeV and NA48 experiments.  For a highly accessible
review of these results, see ref.~\cite{Kleinknecht:2006fu}.

In order to pursue the goal of discovering CP violation the B-meson
systems, two B-factories, one at KEK and one at SLAC, have been built.
We recall here very quickly that these factories 
collide electrons and positrons of differing energies,
which are tuned to produce a `resonance' known as $\Upsilon(4 S)$,
which is a bound state of the $b-\overline{b}$ system.
This resonance decays rapidly into a $B^+-B^-$ pair or $B^0-\bar{B}^0$
pair. Due to the unequal energies of the colliding particles, 
the $\Upsilon(4S)$ resonance, and its decay products, are 
boosted in the laboratory frame. Because of the relativistic
time dilation, the lifetimes of these boosted B-mesons are 
larger than their natural lifetimes. Thus the time
of the decay of each meson could be accurately determined. This
time information plays a crucial role  
in making many of the observations possible.
With the present day luminosities achived by these accelerators,
today there exists a data set consisting of more than a {\it billion}
B-meson pairs which are being studied thoroughly by the detectors 
Belle and BaBar, at these facilities.

As in the case of the K-meson system, CP violation in the B-meson system
was first discovered via indirect CP violation in 2002.  This
discovery was reviewed in this journal in ref.~\cite{Panchapakesan},
in {\it Nature} in ref.~\cite{Peskin2}.  Direct CP
violation was discovered later, in 2004~\cite{directcp}.

The Belle collaboration reports the results obtained from studying a
sample of 535 million B-meson pairs, and looks at a final state 
containing one K-meson and one pion, as well as a final state containing
two pions.  These result from the change of the flavour of the B-meson,
which decays into two lighter mesons.  These decays can be represented
in terms of `Feynman diagrams' where the b-quark which is confined inside
a meson decays through the emission of one of the force carriers of
the weak interaction, namely the $W$ boson.  The decay `amplitude',
the mathematical expression that governs the strength of the
probability of such a transition, also receives
quantum mechanical contributions through virtual quanta that
run around in `loops' which can, for a process of this sort,
be comparable in magnitude to the direct decay amplitude. 
Such diagrams have come to be known as `penguin' diagrams. It is
the quantum mechanical interference of these diagrams that leads 
to the decay rate of $B^+ \to K^+ \pi^0$ being different from that
of $B^- \to K^- \pi^0$ and similarly for the decays $B^0 \to K^+ \pi^-$
and $\bar{B}^0 \to K^- \pi^+$. This difference is the signature of
{\it direct CP violation} and is parametrized in terms of a dimensionless
number known as `CP-asymmetry'.  
Such asymmetries are expected to be comparable for those arising 
from the neutral B-mesons and for the charged B-mesons.  
The discovery that is being reported by Belle as well as BaBar
is that there is a significant mismatch between these asymmetries.  
This is expressed as $7\%$ and $-10\%$ by Belle.  For a discussion
on these points, see ref.~\cite{Peskin}.

Since there are always `strong interaction' effects which are
difficult to estimate, as the quarks are confined within mesons, it
could be that contributions in the standard model which are expected
to be small are actually not so.  Furthermore, the penguins
referred to earlier come in several varieties, of which only one, namely 
the gluon penguin is 
expected to be dominant.  It could so happen that sub-dominant ones
may be enhanced due to (a) our poor understanding of the dynamics, or 
(b) could be receiving contributions from hitherto undiscovered physics.  
This latter possibility is particularly exciting because it points to
a possible new source of CP violation (other than the phase $\delta$
introduced by Kobayashi and Maskawa). 
This has been eloquently referred to as the `song of the electro-weak penguin'
by Peskin in his article.  

We now turn to other interesting data on B-mesons from other 
experiments. The Fermilab experiments CDF and D\O, in the USA have 
recently reported important observations concerning properties of 
so-called $B_s$ mesons. These are a pair of neutral mesons with 
quark content $B_s=\bar{b}s$ and $\bar{B}_s = b\overline{s}$.
In a previous publication~\cite{ACMPRS} 
oscillation phenomena in the $B_s-\overline{B}_s$ system at these
Fermilab experiments soon after their discovery were reviewed. 
We now turn to some of the properties of the $B_s$ meson system that have
been measured subsequently. 

\begin{center}
\begin{figure} \label{basicfig}
\begin{picture}(500,100)(-75,0)
\ArrowLine(-30,100)(60,100)
\ArrowLine(60,0)(-30,0)
\Photon(0,100)(35,50){2}{6}
\ArrowLine(60,80)(35,50)
\ArrowLine(35,50)(60,20)
\Text(-40,100)[]{$b$}
\Text(70,100)[]{$c$}
\Text(0,80)[]{$W^-$}
\Text(70,20)[]{$s$}
\Text(70,80)[]{$\overline{c}$}
\Text(90,90)[]{$J/\psi$}
\Text(90,10)[]{$\phi$}
\Text(-40,0)[]{$\overline{s}$}
\Text(70,0)[]{$\overline{s}$}
\Text(-45,50)[]{$\overline{B}_s$}
\Text(100,50)[]{$,$}
\ArrowLine(220,100)(130,100)
\ArrowLine(130,0)(220,0)
\Photon(160,100)(195,50){2}{6}
\ArrowLine(195,50)(220,80)
\ArrowLine(220,20)(195,50)
\Text(120,100)[]{$\overline{b}$}
\Text(230,100)[]{$\overline{c}$}
\Text(160,80)[]{$W^+$}
\Text(230,20)[]{$\overline{s}$}
\Text(230,80)[]{$c$}
\Text(250,90)[]{$J/\psi$}
\Text(250,10)[]{$\phi$}
\Text(120,0)[]{${s}$}
\Text(230,0)[]{${s}$}
\Text(125,50)[]{${B}_s$}
\end{picture}
\caption{Decay of $B_s$ mesons into $J/\psi\, \phi$}
\end{figure}
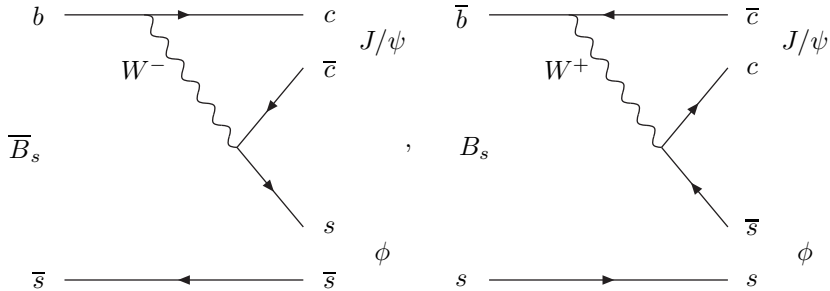
\end{center}

The CDF experiment~\cite{arXiv:0712.2397} and the 
{D\O} experiment~\cite{0802.2255}
considered the particular final
state $J/\psi \phi$. In Fig.~1 we show how the
decays of both $B_s$ as well as $\overline{B}_s$ can 
lead to this final state.  We recall that the $J\psi$ is
a bound state of a charm and anti-charm ({\it viz,} it is
a `charmonium' resonance) and the $\phi$ is a corresponding
resonance with s-quark and its anti-quark. $J/\psi$ is detected
through its decay into $\mu^+ \mu^-$ and $\phi$ through its
decay into $K^+ K^-$. All these particles are relatively long 
lived and hence it is straight forward to measure their momenta
and energies. Thus a full reconstruction of the $B_s/\overline{B}_s$
along with the determination of the CP properties of the final states 
is possible. The difference in the rates $B_s \to J/\psi \phi$ 
and $\overline{B}_s \to J/\psi \phi$ is a measure of the CP violation
in the $B_s$ system. This is predicted to be small in the Kobayashi-Maskawa
model and the experimental results are in accordance. The difference
in the rates between decays into CP even and CP odd final states gives
the lifetime difference between the two $B_s$ mesons. This is predicted
to be measurably large. The measured value from
untagged decays is 
$\Delta \Gamma = 0.076 
\pm 0.060$ ps$^{-1}$, 
which is consistent with zero
(for a review see ref.~\cite{0710.1789}). 
However, the present tagged experiments now yield the number
for this quantity
$\Delta \Gamma = 0.19  
\pm 0.07$ ps$^{-1}$ from D\O, ref.~\cite{0802.2255}.
A useful review of all these experimental results is  
ref.~\cite{0805.2302}.  
On the theoretical front, a case for the possibility of discovery of
new physics based on this data is the analysis of 
ref.~\cite{arXiv:0803.0659}. 

To summarize, in this note we have reviewed some interesting experimental
developments in the sector involving B-mesons and $B_s$ mesons
at a variety of experiments at $e^+ e^-$
colliders and at the Tevatraon, the proton-antiproton collider. 
Having entered a era of precision studies in this sector, it is
clear that many interesting and exciting developments lie ahead,
both experimentally and theoretically, as a possible window to
physics beyond the standard model may open up in these systems.

\end{document}